# ON THE E-ΔE LOW PRESSURE PENTANE FILLED DETECTION MODULE OF DGFRS-2 SETUP

*D. Ibadullayev[a, b, c*], A.N. Polyakov[a], Yu. S. Tsyganov[a], A. A. Voinov[a], M.V. Shumeiko[a]*

[a]*Joint Institute for Nuclear Research, 141980 Dubna, Russian Federation.*

[b]*Institute of Nuclear Physics, 050032 Almaty, Kazakhstan.*

[c]*L.N. Gumilyov Eurasian National University, 010000 Astana, Kazakhstan.*

E-mail: Ibadullayev@jinr.ru

**Abstract**

The paper describes a new E-ΔE detection module which includes a position-sensitive double-sided strip detector (DSSD) and a low-pressure pentane-filled ΔE chamber for detection of Evaporation Residues (ERs) in complete fusion reactions. The goal is to synthesize superheavy isotopes at the new DC-280 (Dubna Cyclotron) cyclotron of the Joint Institute for Nuclear Research (JINR). With the $^{48}$Ca ion beam intensity of about 5-7 pμA (particle micro ampere) and the ΔE counter count rate of approximately $10^4$ sec$^{-1}$, we observe a decrease in the overall registration efficiency of the gas ΔE counter. The paper proposes a solution to this problem: introduction of an additional negatively biased electrode, which results in the stabilization of the ΔE detector operation.

## 1. Introduction

Over the past three decades, the Dubna Gas-Filled Recoil Separator (DGFRS) has been the facility where 5 new elements (from Z = 114 to Z = 118) and more than 50 new super-heavy isotopes were synthesized for the first time [1-2]. In these experiments, we used two gas counters to measure the time of flight (TOF) of evaporation residues (ER) and other charged particles [3-6]. $^{48}$Ca$^{+18}$ ions were accelerated at the FLNR U-400 cyclotron with a beam intensity of ≤ 1 pμA. To synthesize heavier nuclei with Z = 119, 120 in reactions with actinide targets, heavier beams of higher intensities ($^{50}$Ti, $^{54}$Cr, $^{58}$Fe, etc.) are required due to the small cross sections of the ER formation.

## 2. Methodology

In 2020, the improved DGFRS-2 setup was put into operation at the FLNR, online with the new super-intensive cyclotron DC-280. Successful experiments studying the complete fusion reaction induced by heavy ions were performed with beam intensities from 2 to 6.5 pμA [7-10]. Unlike the DGFRS-1 setup [11] designed according to the D-$Q_1$-$Q_2$ scheme (D - Dipole magnet, $Q_1$ – first quadrupole lense, $Q_2$ – second quadrupole lense), the new separator is designed according to a scheme that provides better background clearing, namely $Q_1$-$D_1$-$Q_2$-$Q_3$-$D_2$ ($Q_1$ – first quadrupole lense, $D_1$ – first dipole magnet, $Q_2$ – second quadrupole lense, $Q_3$ – third quadrupole lense, $D_2$ – second dipole magnet).

After passing through magnetic elements of the separator, recoil nuclei (ER) and other background particles enter the detection module. The DGFRS-2 uses a double-sided strip detector (DSSD) 48x128 strip detector (a double-sided silicon strip focal plane detector) measuring 48x220 mm (front strip width - 1 mm, back strip width - 2 mm). The focal plane is surrounded by side strip detectors consisting of 64 strips (each strip is 15 mm wide). A low pressure ΔE gas detector (1.2-1.3 Torr) is also used to register the passage of charged particles before they are implanted into the DSSD. The operating parameters of the gas detector are as follows: cathode - made of 20 μmW(Au) (Gold-coated tungsten) wire, pitch 1 mm; anode - made of 10 μm W(Au) wire, pitch 2 mm, transparency – 91%. A schematic diagram of this module is shown in Figure 1. The entire detection system works as a two-branch system. One branch is a digital system using PIXIE-16 digital modules (Clock speed – 100 MHz), while the other is a CAMAC-based analog system that provides real-time registering of correlated energy-time-coordinate chains of the ER-α type with subsequent switching off of the DC-280 cyclotron beam. This ensures virtually background-free detection of subsequent alpha decays of daughter nuclei [4-6, 12]. Figure 2 shows a block diagram of a PC-based data-acquisition system (C++ Builder software, see [13-15]).

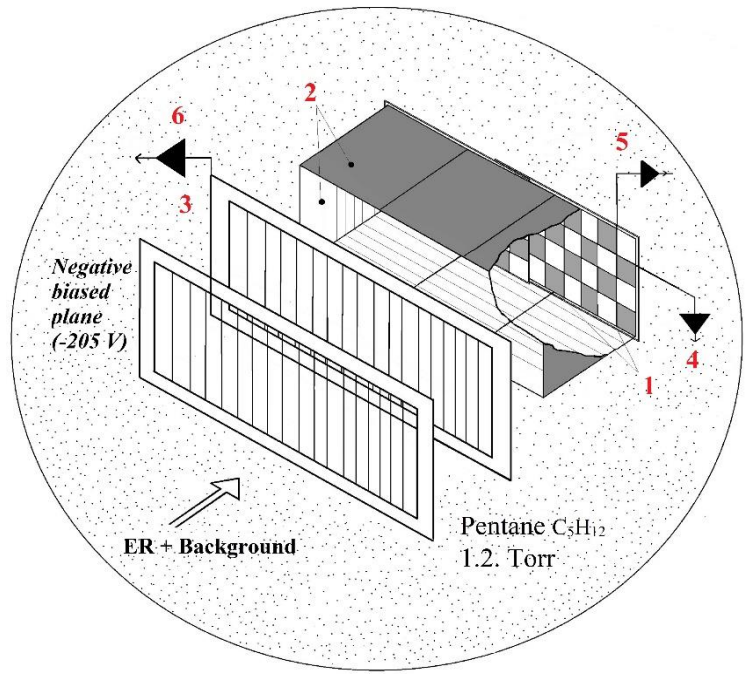

Figure 1. Detection Module of the DGFRS-2 setup. In figure 1 – DSSD Detectors; 2 – Side Detectors; 3 – MWPC-ΔE; 4-6 – Front, Back, ΔE Preamplifiers respectively.

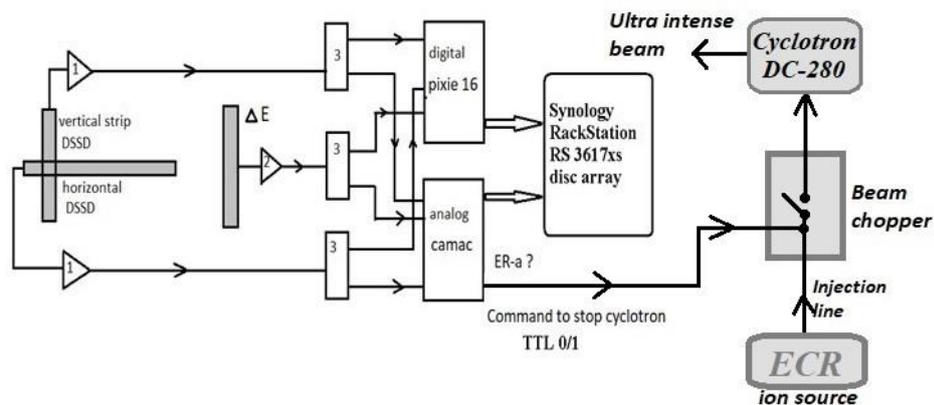

Figure 2. The schematic of the PC-based data acquisition system. Only one vertical and one horizontal strips are shown. 1– MPR-64Mesitec charge sensitive preamplifier. 2 – PUPS-1 charge sensitive preamplifier (custom made). 3 – Preamplifier signal splitter (custom, FLNR design).

These experiments use modern commercially available Pixie digital boards to form the main data block for processing with minimal dead time (on the order of ~ 100 ns, instead of 25 μs (CAMAC)). But, we note as a disadvantage of this system, the fact that it operates through large internal buffers, which excludes (at least at the level of a simple solution) the use of the author's method of "active correlations" [4, 5, 13, 15]. At the same time, it is the analog spectrometer that serves this purpose. This makes it possible to radically suppress the background of charged particles associated with the operation of the powerful DC-280 accelerator. An auxiliary function for the analog spectrometer is the process of data duplication and the on-line monitoring of the ΔE detector efficiency parameter.

3. **Results and Discussion**

Unlike [4-6], we changed the operating mode of the gas module for the following reasons (see Fig. 1):

– In some cases, when the module count rate exceeds the range of $7\text{-}8\cdot10^3$ to $2\cdot10^4$, the detection efficiency drops to 40-70% for an extended period of time.

– Sometimes this value may spontaneously drop to zero.

The solution to this problem was the use of the "start" chamber as an electron deflection plane to provide additional injection of electrons into the "stop" counter, resulting in easier dissipation of the relatively slow positive ion cloud near the "stop" counter (see, e.g., [12]).

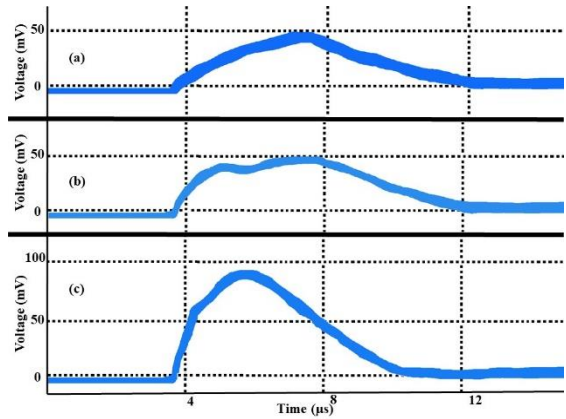

Figure 3. Heavy ion ΔE signal amplitude transformation with deflecting plane voltage: (a) 0 V, (b) -40 V, (c) -205 V.

In Figure 3, the process is illustrated by representing the ΔE output signal from a charge-sensitive preamplifier with different deflection plane voltages. Figure 3a shows the ΔE output signal at a voltage of 0 V without injection of electrons into the "stop" counter. Figure 3b shows the output signal at -40V, where minimal injection occurs and a mixed state is observed. In Figure 3c, at an output signal voltage of -205 V, maximum injection of electrons into the "stop" counter occurs, and we observe a 2-fold increase in the signal value. A further increase in the value of the negative bias up to -270 V gives virtually no increase in the signal amplitude.

Figures 4a and 4b show the registered ΔE signal values against typical background signals for the reactions $^{242}$Pu + $^{48}$Ca→Fl* from the analog electronics branch and $^{232}$Th + $^{48}$Ca→Ds* from the digital electronics branch, respectively [8, 10].

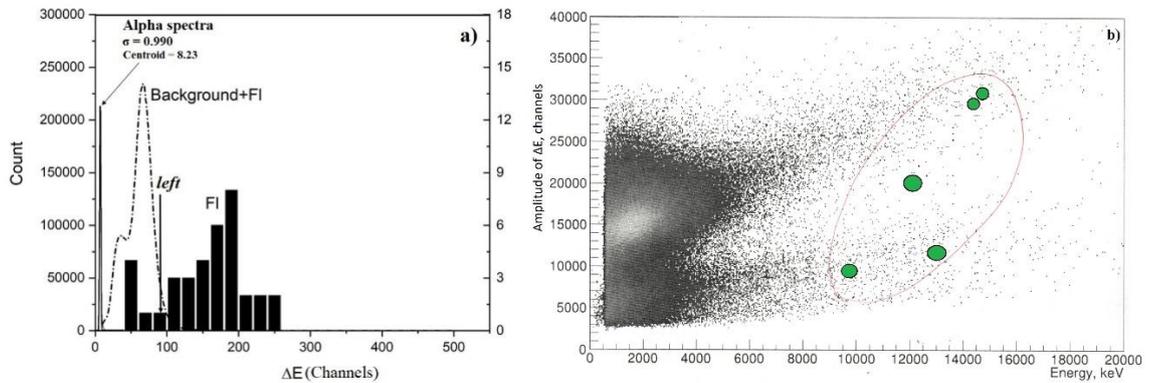

Figure 4a. ΔE signals of background (dash-dotted line), Fl$^{114}$ (black histogramm) registered in experiment $^{242}$Pu + $^{48}$Ca and alpha spectra from 5.5 MeV external source (solid line) (*analog electronics branch*). b. 2d spectra of the dependence of ΔE signals on energy and amplitude of background plus evaporation residues registered in $^{232}$Th + $^{48}$Ca experiment, Ds signals registered in experiment are shown with green circles (*digital electronics branch*).

Note that in both cases the study products are separated a little on the ΔE scale. The registered energy amplitudes of the evaporation residues are in good agreement with the calculated method presented in [12]. Almost the same results were experimentally obtained in 2021 in the reaction $^{243}$Am + $^{48}$Ca → Mc* and in 2023 in the reaction $^{238}$U + $^{54}$Cr → Lv* at the DGFRS-2 setup. It should be noted that amplitude analysis can provide an additional cleaning factor ξ of approximately ~15 for the reaction $^{242}$Pu + $^{48}$Ca → Fl*. We define the ξ factor as:

$$\xi = \frac{\sum_{0}^{End}\Delta E(i)}{\sum_{left}^{End}\Delta E(i)},$$

And the "left" channel corresponds to approximately 82% of the events of the sought nuclide Fl located to the right relative to the "left" channel (see Fig. 4a).

4. **Summary**

The new operating mode of the ΔE chamber was studied in long-term experiments at the DGFRS-2 setup operating with super-intense heavy ion beams of the FLNR (JINR) DC-280 cyclotron. Partial separation of the amplitudes of registered evaporation residues (ER's) in the spectrum of ΔE values was shown as compared to background products. In addition, a method has been found to minimize the negative effect of a decrease in the detection efficiency of a gas ΔE detector at ultra-high intensities of a heavy ion beam. We plan to use this operating mode in the coming experiments, and also to study the specific features of this mode both for various complete fusion reactions and for various operating parameters of the low-pressure gas module itself.

5. **Supplement 1. The present status of on-line monitoring of the ΔE detector efficiency parameter.**

During the writing of this paper, the experiment $^{238}U + {}^{54}Cr \rightarrow Lv^*$ was successfully performed at the DGFRS-2 setup. On–line monitoring of the efficiency value was carried out for three energy intervals measured with a DSSD detector, specifically: 2-4, 20-30, 40-250 MeV. Typical results are presented in Table 1 for eight operational files. In each case, the efficiency values were defined as: $\varepsilon = \frac{N_{\Delta E}}{N_{ALL}}$.

Here, $N_{\Delta E}$ is a number of events with ΔE > 0, and $N_{ALL}$ – total number of events within given energy interval.

Table 1. Measured values of the ΔE detector efficiency.

| File № | 838 | 843 | 845 | 840 | 833 | 889 | 948 | 930 | mean |
|---|---|---|---|---|---|---|---|---|---|
| Interval1 | 0.998 | 0.999 | 0.999 | 0.999 | 0.999 | 0.995 | 0.975 | 0.992 | 0.9945 |
| Interval2 | 1.000 | 1.000 | 1.000 | 1.000 | 0.991 | 1.000 | 1.000 | 0.997 | 0.9985 |
| Interval3 | 1.000 | 0.986 | 1.000 | 1.000 | 0.977 | 1.000 | 1.000 | 0.990 | 0.9941 |

6. **Supplement 2. Registered amplitudes of superheavy recoils**

Below, spectra of ER registered energy measured with DSSD detector are shown in the Fig.5a-c. Dotted lines (a) correspond to PC simulations reported in [12]. Note, that about 6% of events for Mc (5a) recoilas are outside calculated area.

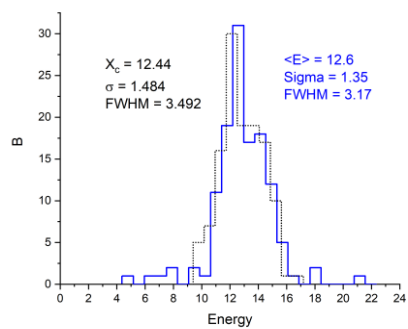

Fig.5a Measured and calculated (dot line) spectra of Mc ER's. Reaction $^{243}Am+^{48}Ca\rightarrow Mc^*$

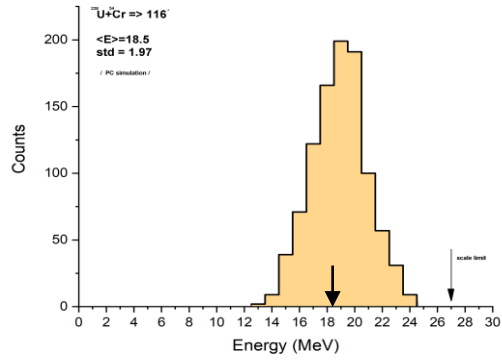

Fig 5b. PC simulated spectrum of Lv recoils. One registered event is shown by…Reaction $^{238}U+^{54}Cr \rightarrow Lv^*$.

*( one measured event of Lv is shown by arrow)*

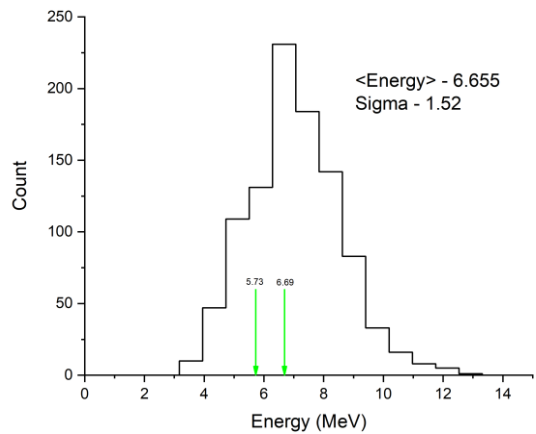

Fig. 5c PCsimulated spectrum of Ds ER's. Two measured events are shown by errors. Reaction $^{238}U+^{40}Ar \rightarrow Ds^*$.

7. **Supplement 3 Typical dependence of DSSD leakage current against dose value.**

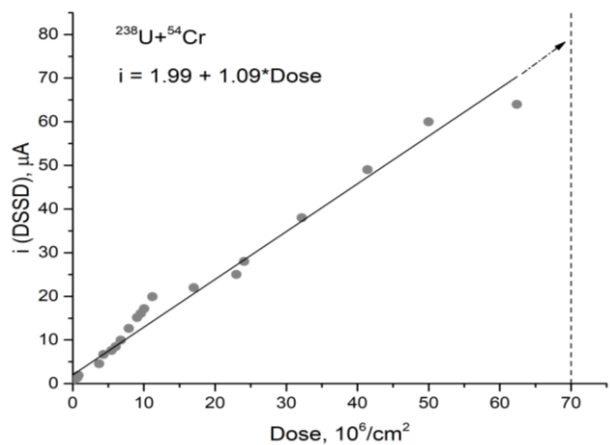

Fig.6 Dependense of leakage current value agains heavy ion dose ( mostly ~6 Mev U target like products)

Beam dose value is calculated due to: $Dose = \frac{N}{A_{eff}}$, where N- number of ions coming onto DSSD and $A_{eff}$ is effective area (see Fig.7, yellow area) for 68% of particles level for $^{238}U+^{54}Cr \rightarrow Lv^*$ reaction.

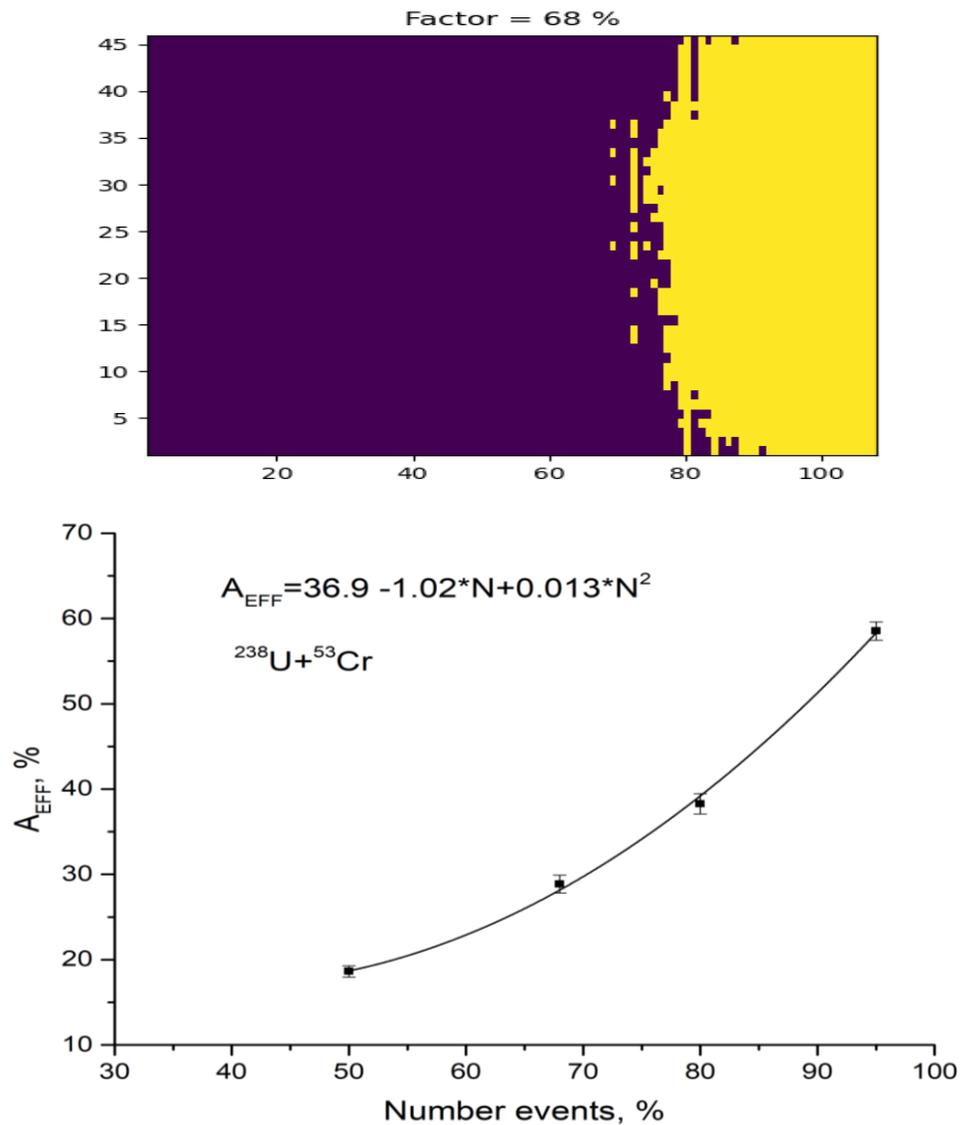

Fig.7 Dependense of effective area against level of content